\documentclass[twocolumn]{aastex631}
\usepackage{changepage}
\usepackage[graphicx]{realboxes}
\usepackage{lineno}

\received{XXX}
\accepted{XXX}

\submitjournal{ApJL}

\shorttitle{PHANGS-JWST Embedded Clusters}
\shortauthors{PHANGS Collaboration}

\begin{document}

\title{PHANGS-JWST First Results: Dust embedded star clusters in NGC~7496\\ selected via 3.3$\mu$m PAH emission}

\author[0000-0002-0579-6613]{M. Jimena Rodríguez}
\affiliation{Steward Observatory, University of Arizona, 933 N Cherry Ave,Tucson, AZ 85721, USA}
\affiliation{Instituto de Astrofísica de La Plata, CONICET--UNLP, Paseo del Bosque S/N, B1900FWA La Plata, Argentina }

\author[0000-0003-0946-6176]{Janice C. Lee}
\affiliation{Gemini Observatory/NSF’s NOIRLab, 950 N. Cherry Avenue, Tucson, AZ, USA}
\affiliation{Steward Observatory, University of Arizona, 933 N Cherry Ave,Tucson, AZ 85721, USA}

\author[0000-0002-3784-7032]{B. C. Whitmore}
\affiliation{Space Telescope Science Institute, 3700 San Martin Drive, Baltimore, MD 21218, USA}

\author[0000-0002-8528-7340]{David A. Thilker}
\affiliation{Department of Physics and Astronomy, The Johns Hopkins University, Baltimore, MD 21218 USA}

\author[0000-0001-6038-9511]{Daniel Maschmann}
\affiliation{Steward Observatory, University of Arizona, 933 N Cherry Ave,Tucson, AZ 85721, USA}
\affiliation{Sorbonne {Universit\'e},
LERMA, Observatoire de Paris, PSL university, CNRS, F-75014, Paris, France}

\author[0000-0003-0085-4623]{Rupali~Chandar}
\affiliation{Ritter Astrophysical Research Center, University of Toledo, Toledo, OH 43606, USA}

\author[0000-0003-1943-723X]{Sinan Deger}
\affiliation{The Oskar Klein Centre for Cosmoparticle Physics, Department of Physics, Stockholm University, AlbaNova, Stockholm, SE-106 91, Sweden}

\author[0000-0003-0946-6176]{Médéric~Boquien}
\affiliation{Centro de Astronomía (CITEVA), Universidad de Antofagasta, Avenida Angamos 601, Antofagasta, Chile}

\author[0000-0002-5782-9093]{Daniel~A.~Dale}
\affiliation{Department of Physics and Astronomy, University of Wyoming, Laramie, WY 82071, USA}

\author[0000-0003-3917-6460]{Kirsten L. Larson}
\affiliation{AURA for the European Space Agency (ESA), Space Telescope Science Institute, 3700 San Martin Drive, Baltimore, MD 21218, USA}

\author[0000-0002-0012-2142]{Thomas G. Williams}
\affiliation{Sub-department of Astrophysics, Department of Physics, University of Oxford, Keble Road, Oxford OX1 3RH, UK}
\affiliation{Max-Planck-Institut f\"ur Astronomie, K\"onigstuhl 17, D-69117 Heidelberg, Germany}

\author[0000-0003-4770-688X]{Hwihyun~Kim}
\affiliation{Gemini Observatory/NSF’s NOIRLab, 950 N. Cherry Avenue, Tucson, AZ, USA}

\author[0000-0002-3933-7677]{Eva Schinnerer}
\affiliation{Max-Planck-Institut f\"ur Astronomie, K\"onigstuhl 17, D-69117 Heidelberg, Germany}

\author[0000-0002-5204-2259]{Erik Rosolowsky}
\affiliation{Department of Physics, University of Alberta, Edmonton, Alberta, T6G 2E1, Canada}

\author[0000-0002-2545-1700]{Adam~K.~Leroy}
\affiliation{Department of Astronomy, The Ohio State University, 140 West 18th Avenue, Columbus, Ohio 43210, USA}
\affiliation{Center for Cosmology and Astroparticle Physics, 191 West Woodruff Avenue, Columbus, OH 43210, USA}

\author[0000-0002-6155-7166]{Eric Emsellem}
\affiliation{European Southern Observatory, Karl-Schwarzschild-Stra{\ss}e 2, 85748 Garching, Germany}
\affiliation{Univ Lyon, Univ Lyon1, ENS de Lyon, CNRS, Centre de Recherche Astrophysique de Lyon UMR5574, F-69230 Saint-Genis-Laval France}

\author[0000-0002-4378-8534]{Karin M. Sandstrom}
\affiliation{Center for Astrophysics \& Space Sciences, University of California, San Diego, 9500 Gilman Drive, San Diego, CA 92093, USA}

\author[0000-0002-8804-0212]{J.~M.~Diederik~Kruijssen}
\affiliation{Cosmic Origins Of Life (COOL) Research DAO, coolresearch.io}

\author[0000-0002-3247-5321]{Kathryn~Grasha}
\affiliation{Research School of Astronomy and Astrophysics, Australian National University, Canberra, ACT 2611, Australia}   
\affiliation{ARC Centre of Excellence for All Sky Astrophysics in 3 Dimensions (ASTRO 3D), Australia}

\author[0000-0002-7365-5791]{Elizabeth~J.~Watkins}
\affiliation{Astronomisches Rechen-Institut, Zentrum f\"{u}r Astronomie der Universit\"{a}t Heidelberg, M\"{o}nchhofstra\ss e 12-14, D-69120 Heidelberg, Germany}

\author[0000-0003-0410-4504]{Ashley.~T.~Barnes}
\affiliation{Argelander-Institut f\"{u}r Astronomie, Universit\"{a}t Bonn, Auf dem H\"{u}gel 71, 53121, Bonn, Germany}

\author[0000-0001-6113-6241]{Mattia C. Sormani}
\affiliation{Universit\"{a}t Heidelberg, Zentrum f\"{u}r Astronomie, Institut f\"{u}r Theoretische Astrophysik, Albert-Ueberle-Stra{\ss}e 2, D-69120 Heidelberg, Germany}

\author[0000-0002-0432-6847]{Jaeyeon Kim}
\affiliation{Universit\"{a}t Heidelberg, Zentrum f\"{u}r Astronomie, Institut f\"{u}r Theoretische Astrophysik, Albert-Ueberle-Stra{\ss}e 2, D-69120 Heidelberg, Germany}

\author[0000-0002-5259-2314]{Gagandeep S. Anand}
\affiliation{Space Telescope Science Institute, 3700 San Martin Drive, Baltimore, MD 21218, USA}

\author[0000-0002-5635-5180]{M\'elanie Chevance}
\affiliation{Universit\"{a}t Heidelberg, Zentrum f\"{u}r Astronomie, Institut f\"{u}r Theoretische Astrophysik, Albert-Ueberle-Stra{\ss}e 2, D-69120 Heidelberg, Germany}
\affiliation{Cosmic Origins Of Life (COOL) Research DAO, coolresearch.io}

\author[0000-0003-0166-9745]{F. Bigiel}
\affiliation{Argelander-Institut f\"ur Astronomie, Universit\"at Bonn, Auf dem H\"ugel 71, 53121 Bonn, Germany}

\author[0000-0002-0560-3172]{Ralf S.\ Klessen}
\affiliation{Universit\"{a}t Heidelberg, Zentrum f\"{u}r Astronomie, Institut f\"{u}r Theoretische Astrophysik, Albert-Ueberle-Stra{\ss}e 2, D-69120 Heidelberg, Germany}
\affiliation{Universit\"{a}t Heidelberg, Interdisziplin\"{a}res Zentrum f\"{u}r Wissenschaftliches Rechnen, Im Neuenheimer Feld 205, D-69120 Heidelberg, Germany}

\author[0000-0002-8806-6308]{Hamid Hassani}
\affiliation{Department of Physics, University of Alberta, Edmonton, Alberta, T6G 2E1, Canada}

\author[0000-0001-9773-7479]{Daizhong Liu}
\affiliation{Max-Planck-Institut f\"ur Extraterrestrische Physik (MPE), Giessenbachstr. 1, D-85748 Garching, Germany}

\author[0000-0001-5310-467X]{Christopher M. Faesi}
\affiliation{University of Connecticut, Department of Physics, 196A  Auditorium Road, Unit 3046, Storrs, CT, 06269}

\author[0000-0001-5301-1326]{Yixian Cao}
\affiliation{Max-Planck-Institut f\"ur Extraterrestrische Physik (MPE), Giessenbachstr. 1, D-85748 Garching, Germany}

\author[0000-0002-2545-5752]{Francesco Belfiore}
\affiliation{INAF — Arcetri Astrophysical Observatory, Largo E. Fermi 5, I-50125, Florence, Italy}

\author[0000-0002-0873-5744]{Ismael Pessa}
\affiliation{Max-Planck-Institut f\"ur Astronomie, K\"onigstuhl 17, D-69117 Heidelberg, Germany}
\affiliation{Leibniz-Institut f\"{u}r Astrophysik Potsdam (AIP), An der Sternwarte 16, 14482 Potsdam, Germany}

\author[0000-0001-6551-3091]{Kathryn~Kreckel}
\affiliation{Astronomisches Rechen-Institut, Zentrum f\"{u}r Astronomie der Universit\"{a}t Heidelberg, M\"{o}nchhofstra\ss e 12-14, D-69120 Heidelberg, Germany}

\author[0000-0002-9768-0246]{Brent Groves}
\affiliation{International Centre for Radio Astronomy Research, University of Western Australia, 7 Fairway, Crawley, 6009 WA, Australia}

\author[0000-0003-3061-6546]{Jérôme Pety}
\affiliation{IRAM, 300 rue de la Piscine, 38400 Saint Martin d'H\`eres, France}
\affiliation{LERMA, Observatoire de Paris, PSL Research University, CNRS, Sorbonne Universit\'es, 75014 Paris}

\author[0000-0002-4663-6827]{R\'emy Indebetouw}
\affiliation{National Radio Astronomy Observatory, 520 Edgemont Rd., Charlottesville, VA 22903}
\affiliation{University of Virginia, 530 McCormick Rd., Charlottesville, VA 22904}

\author[0000-0002-4755-118X]{Oleg V.~Egorov}
\affiliation{Astronomisches Rechen-Institut, Zentrum f\"{u}r Astronomie der Universit\"{a}t Heidelberg, M\"{o}nchhofstra\ss e 12-14, D-69120 Heidelberg, Germany}

\author[0000-0003-4218-3944]{Guillermo A. Blanc}
\affiliation{The Observatories of the Carnegie Institution for Science, 813 Santa Barbara St., Pasadena, CA, USA}
\affiliation{Departamento de Astronom\'{i}a, Universidad de Chile, Camino del Observatorio 1515, Las Condes, Santiago, Chile}

\author[0000-0002-2501-9328]{Toshiki Saito}
\affiliation{National Astronomical Observatory of Japan, 2-21-1 Osawa, Mitaka, Tokyo, 181-8588, Japan}

\author[0000-0002-9181-1161]{Annie~Hughes}
\affiliation{IRAP, Universit\'e de Toulouse, CNRS, CNES, UPS, (Toulouse), France} 




\begin{abstract}

The earliest stages of star formation occur enshrouded in dust 
and are not observable in the optical.  
Here we leverage the extraordinary new high-resolution infrared imaging from JWST to begin the study of dust-embedded star clusters in nearby galaxies throughout the local volume.  We present a 
technique for identifying dust-embedded clusters in NGC~7496 (18.7~Mpc), the first galaxy to be observed by the PHANGS-JWST Cycle 1 Treasury Survey. 
We select sources that have strong 3.3$\mu$m PAH emission based on a $\rm F300M-F335M$ color excess, and identify 67 candidate embedded clusters.  Only eight of these are found in the PHANGS-HST optically-selected cluster catalog and all are young (six have SED-fit ages of $\sim1$~Myr).  We find that this sample of embedded cluster candidates may significantly increase the census of young clusters in NGC~7496 from the PHANGS-HST catalog -- the number of clusters younger than $\sim$2 Myr could be increased by a factor of two.  Candidates are preferentially located in dust lanes, and are coincident with peaks in PHANGS-ALMA CO (2-1) 
maps. We take a first look at concentration indices, luminosity functions, SEDs spanning from 2700\AA~to 21$\mu$m, and stellar masses (estimated to be between $\sim10^4-10^5 M_{\odot}$).  The methods tested here provide a basis for future work to derive accurate constraints on the physical properties of embedded clusters, characterize the completeness of cluster samples, and expand analysis to all 19 galaxies in the PHANGS-JWST sample, which will enable basic unsolved problems in star formation and cluster evolution to be addressed.

\end{abstract}



\section{Introduction} \label{sec:intro}

Star clusters are formed in the densest parts of giant molecular clouds \citep[][]{lada03,2008ApJ...672.1006E,kruijssen12}. During their formation and early evolution they remain surrounded by dust and molecular gas causing them to be heavily obscured at optical wavelengths, and often only detectable at infrared and longer wavelengths.  Until now, such young objects were mostly identified using observations of the radio continuum \citep{2003AJ....126..101J} or dust continuum in combination with molecular-gas lines \citep{2018ApJ...869..126L}. 
Infrared studies of the star clusters inside such nurseries has so far mainly occurred in the Milky Way, Magellanic Clouds, and Local Group \citep[e.g.][]{2019A&A...622A.149G}. 

Since most stars form in clustered environments \citep{lada03}, embedded clusters represent a very early, critical stage in the star formation process. Therefore,
the study of these systems is essential for understanding the mechanisms that drive star formation overall, and the properties and evolution of star cluster populations in galaxies.

Previous studies suggest that star clusters remain in this embedded phase within their natal cloud for only a few Myr \citep[$\sim$ 2-5~Myr,  e.g.][Kim et al 2022, Whitmore et al. 2022, this volume]{whitmore14,2017A&A...601A.146C,2021ApJ...909..121M,2018MNRAS.481.1016G,grasha19,2021MNRAS.504..487K}.  However it has not been possible to obtain complete samples of embedded clusters beyond the Local Group to constrain these durations more directly, as the angular resolution of previous infrared facilities could not resolve the parsec scales required to identify clusters at larger distances.

The powerful new infrared capabilities of JWST enable embedded young star clusters throughout nearby galaxies to be identified and studied statistically.  Complete samples of embedded clusters from PHANGS-JWST, combined with the PHANGS-HST, ALMA, and MUSE datasets, will allow basic open issues in star and cluster formation to be addressed, such as fraction of star formation that occurs in clusters (cluster formation efficiency), timescales for the clearing of the natal dust and gas, the conditions that lead to cluster formation, and the dependence of these parameters on galactic environment.

In this pilot study, we identify candidate embedded clusters in NGC~7496 using PHANGS-JWST NIRCam and PHANGS-HST UV-optical imaging.  We find that selection based on an excess in the $\rm F300M-F335M$ color, indicating PAH $3.3\,\mu$m emission, effectively identifies the youngest dusty star clusters.  We take a first look at their properties, and compare with the census of optically-selected clusters in the PHANGS-HST catalog.
The rest of this paper is organized as follows.  In Section~2, we summarize the observations of NGC~7496 taken as part of the PHANGS-JWST Treasury program.  In Section~3 we identify a small set of prototype embedded clusters with strong PAH $3.3\,\mu$m emission, and use their measured properties to define selection criteria to identify a larger smaple of candidate young, embedded clusters.  In Section~4 we estimate the masses of the newly identified clusters and take a first look at the their luminosity functions at 2 and 3 $\mu$m.  We compare this new embedded cluster sample with the optically-selected PHANGS-HST catalog in Section~5, and also examine other properties of the sources, including concentration indices, and UV-IR SEDs, which also incorporates photometry from PHANGS-JWST MIRI imaging. We summarize our main conclusions and outline future work in Section~6.


\section{Data} \label{sec:Data}
The PHANGS-JWST Cycle 1 Treasury program observed its first target, the dusty barred spiral galaxy NGC~7496 located at 18.7~Mpc \citep{2017ApJ...850..207S, 2020AJ....159...67K, 2021MNRAS.501.3621A}, on 6 July 2022. The data were released (simultaneously to the team and public) on 14 July 2022. 
All 19 galaxies in the PHANGS-JWST sample will eventually be imaged with NIRCam and MIRI from 2.0$\micron$ to 21$\micron$, using 8 filters (NIRCam: F200W, F300M, F335M and F360M; MIRI: F770W, F1000W, F1130W and F2100W). Full details of the observing strategy and data reduction are given in the PHANGS-JWST survey description paper in this Volume (Lee et al. 2022).

In this analysis, we examine the JWST data together with UV-optical imaging in 5 filters from the PHANGS-HST survey \citep[][F275W, F336W, F438W, F555W and F814W]{phangs-hst}. 
We compare our dust embedded clusters with the sample from the PHANGS-HST cluster catalog \citep{phangs-hst,thilker21,turner21,deger22}, which 
provides age and stellar-mass estimates from spectral-energy distribution (SED) fitting in NGC~7496. 
Here, we consider the 263 clusters in the HST catalog which have passed human visual inspection and are classified as class 1 (symmetric compact cluster), and class 2 \citep[asymmetric compact cluster;][]{whitmore21,deger22}. This PHANGS-HST sample spans a wide range of properties, with ages between 1~Myr and 13.7~Gyr, and masses between 500$M_\odot$ and 4.6 $\times 10^6 M_\odot$ with a median of 1.5 $\times 10^4 M_\odot$.  Of these, 131 are 10 Myr and younger, with masses between 600$M_\odot$ and 4.3 $\times 10^5 M_\odot$ with a median of 8000 M$_\odot$.
%



\section{Methodology} \label{subsec:Methodology}

\subsection{Identification of embedded cluster prototypes}
\label{subsec:prototypes}

\begin{figure*}
\centering
\includegraphics[width=1.6\columnwidth]{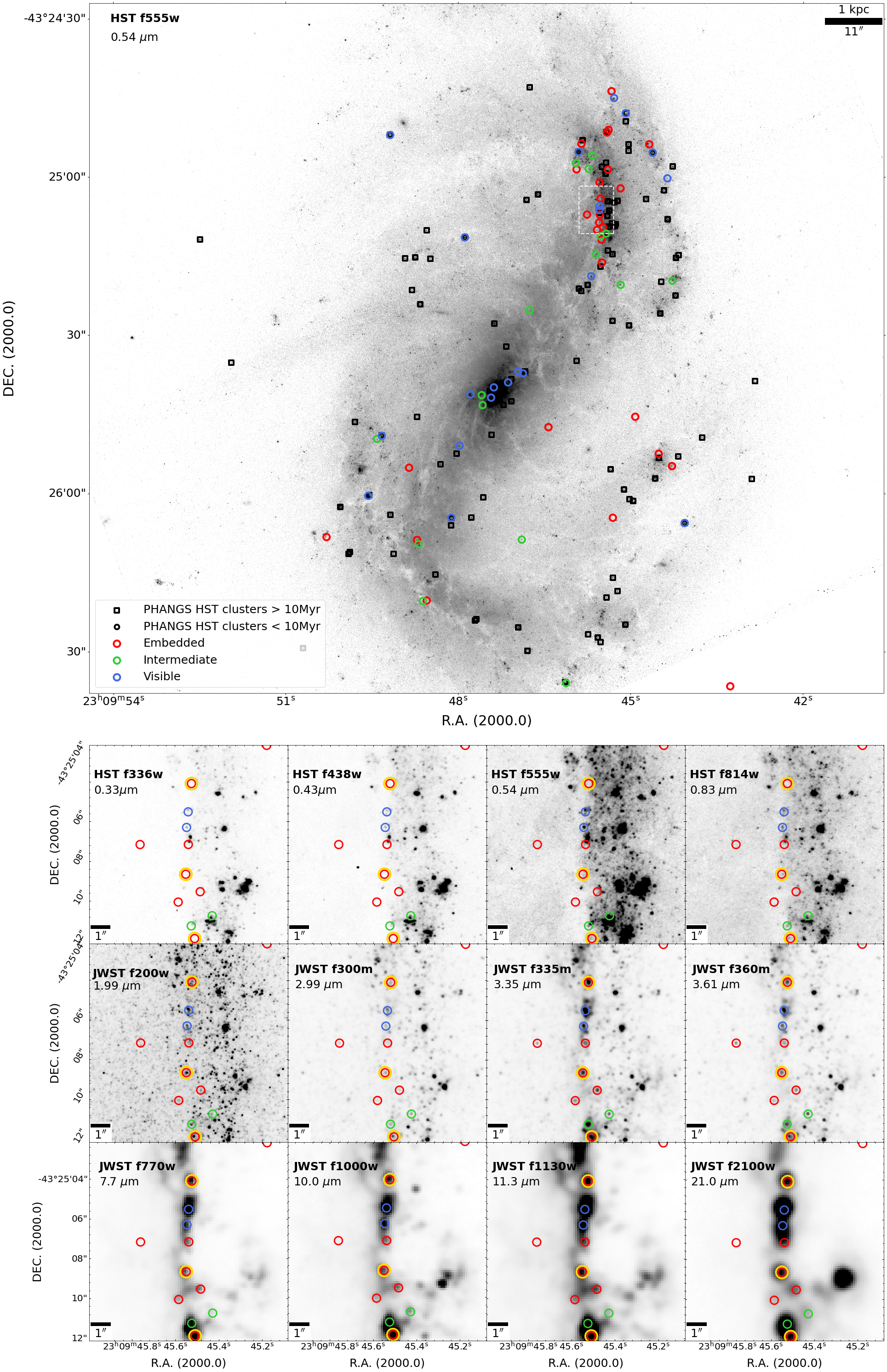}
\caption{
The upper panel shows the PHANGS-HST WFC3 $\rm F555M$ image of NGC~7496. 
The sample of 67 PHANGS-JWST 3.3 $\mu$m selected star cluster candidates are plotted with different colors according to their categories: embedded (red), partially embedded (green) and visible (blue).
The locations of optically-selected clusters from the PHANGS-HST catalog are also shown, and distinguished by SED-fit age (black squares for $>10$ Myr and black circles for $<10$~Myr). Only clusters brighter than $\rm F300M < 25$ are included.
The bottom panels show images in the HST-WFC3, JWST-NIRCam, and JWST-MIRI filters, for the $10" \times 10"$ region indicated by the white box in the upper panel. The central wavelength of each filter is indicated in the panel label.
The colored circles are the same as in the upper panel. Clusters marked by an additional yellow circle are part of our prototype sample (see text for details).
The comparison of this region in the optical HST images and IR from JWST, clearly shows that the candidates fall along a dust lane, largely obscured in the HST bands.
\label{fig:galaxy+reg}}
\end{figure*}

To begin our studies of dust embedded, young star clusters with JWST, we visually  inspect the images to identify a small sample of sources that are obvious candidates for dusty clusters and can serve as prototypes to guide our analysis.
We look for bright compact sources in the NIRCam $\rm F335M$ filter that also appear in at least one other PHANGS-JWST NIRCam image ($\rm F200W$, $\rm F300M$, or $\rm F360M$), but are either faint or not visible by eye in the PHANGS-HST UV-optical images (A$_V \gtrsim$ 10).
 We choose compact objects that are slightly more extended than point sources 
(i.e., FWHM~$\gtrsim$~1.8 pixels $\sim$ 10~pc), 
following a process similar to that for optical star clusters in PHANGS-HST imaging \citep{phangs-hst, thilker21, whitmore21, deger22}.  We choose to start our inspection with the NIRCam $F335M$ image because it is the highest resolution image obtained by PHANGS-JWST that captures PAH emission and hot dust\footnote{We do not use continuum subtracted F335M images for visual inspection to avoid issues with residuals on the small spatial scales of interest here.}. We identify an initial set of 12 prototypical clusters. The bottom panels of  Fig.~\ref{fig:galaxy+reg} show examples of these prototypes marked in yellow. Aperture photometry is performed to examine the properties of these objects and to develop selection criteria to select embedded cluster candidates, as discussed further in the next section. 

\subsection{Source detection and photometry}
\label{sec:idphotometry}
To obtain positions and photometry for our set of 12 prototype embedded clusters, as well as to search for embedded clusters in a more systematic way,  we use routines in the {\sc Photutils} \citep{larry_bradley_2022_6825092} Astropy package for photometry.
First, we use the peak-finding algorithm {\it find\_peaks} to detect sources in the F335M image.
The level of the background varies significantly across the image. So, in order to estimate local background levels, we use the function {\it SExtractorBackground} to build a 2D model of the background. Then, we search for local maxima in square regions that are 11~px on a side. 


To derive accurate source positions, we use the function {\it centroid\_quadratic}, which determines the centroid by fitting 2D quadratic polynomials. 
We extract photometry from a circular aperture with radius of 0\farcs093 (3 NIRCam short-wavelength pixels) which encloses more than 50\% of the energy for a point source (0\farcs073) in the F335W filter.
We subtract the background computed from an annulus between 0\farcs217 and 0\farcs31 away from the source center. The units of the NIRCam imaing are MJy/steradian, so we multiply by the pixel area in steradians, given by the parameter {\it PIXAR\_SR} in the header of the image, to obtain flux densities in MJy.
We then convert the measured flux densities into the AB magnitudes using the usual expression: {$\\ \displaystyle m_{\text{AB}}= -2.5\log _{10}\left({\frac {f_{\nu }}{3\,631 \times 10^{-6}{\text{ MJy}}}}\right).$}\\
For this initial analysis, we forgo aperture corrections to the photometry to calculate total source fluxes.  The measurement of accurate total fluxes for individual clusters across the PHANGS-JWST filter set will be subject of dedicated future work to address the challenges presented by the 10-fold decrease in resolution from 2-21$\mu$m.
Nevertheless, our sample selection strategy which is primarily based on the F300M-F335M color (described in Sect.~\ref{subsec:Methodology}) should not be greatly affected since the resolution in these adjacent bands is quite similar, with a difference in the PSF FWHM of only 0\farcs011.

\subsection{Selection criteria for embedded cluster candidates}
\label{subsect:selectioncriteria}
\begin{figure*}
\centering
\includegraphics[width=2\columnwidth]
{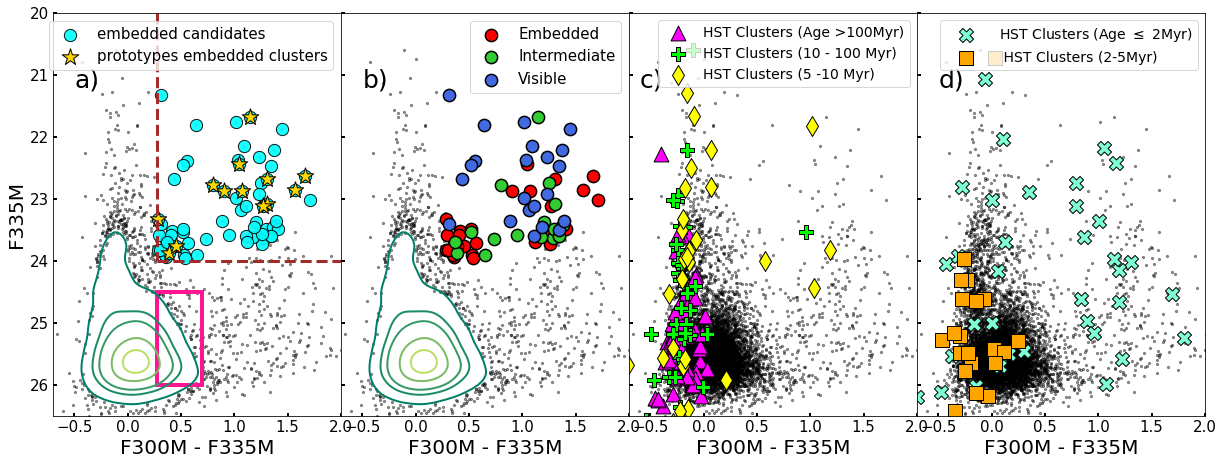}
\caption{Color-magnitude diagram $\rm F335M$ vs. $\rm F300M-F335M$ for NGC~7496. Magnitudes are in the AB system. In all panels the black dots are $\rm F335M$ sources detected with Photutils {\it find\_peaks}. Density contours for the densest parts of this diagram are shown in panels {\it a} and {\it b}. In panel {\it a)} yellow stars show the 12 prototype embedded clusters selected in our initial visual inspection. Based on the location in the CMD of these candidates, we select 67 embedded candidates (cyan circles). The dashed lines show the selection criteria: $\rm F300M-F335M >$0.3 and F335M$<$24. The magenta box indicates a region where scatter due to the uncertainties in the color is large, and is chosen for the purposes of illustrating the spatial distribution in Fig.~\ref{Fig:color_CO_alpha}.
Panel {\it b)} shows the same extended sample of 67 candidates but differentiated by category: embedded objects (red), intermediate or partially embedded (green), and visible or exposed (blue). 
Panels {\it c} and {\it d} show for comparison the location in the CMD sources from the PHANGS-HST optical cluster catalog.
Panel {\it c)} show clusters older than 100~Myr (magenta). As expected, these clusters fall around $\rm F300M-F335M=0$, indicating that these old sources do not have PAH emission. This panel also shows clusters with SED-fit age estimates between 10-100 Myr (green crosses) and 5-10 Myr (yellow diamonds), which also appear around $\rm F300M-F335M=0$. In contrast, about half of the clusters younger than 2~Myr shown in panel {\it d)} (small cyan crosses) are PAH emitters. Panel {\it d)} also shows clusters between 2 and 5~Myr (orange squares), which mostly have little measured PAH emission.} 
\label{fig:cmd}
\end{figure*}

In Figure~\ref{fig:cmd} we show color-magnitude diagrams (CMDs) based on $\rm F335M$ and $\rm F300M$ photometry; the former captures flux from PAH 3.3$\mu$m feature, while the latter primarily probes stellar and dust continuum (Lee et al. 2022; Sandstrom et al 2022, Chastenet et al 2022, this Volume). Panel {\it a} in Figure~\ref{fig:cmd} shows the 12 visually identified prototype embedded clusters (yellow stars) together with the sample of $\rm F335M$ sources detected using Photutils {\it find\_peaks} (black dots).
This latter sample is comprised of a mixture of sources including stellar clusters of different ages, MS, RGB and AGB stars and also background galaxies.

These CMDs are analogous to those commonly used to select emission-line sources in extra-galactic narrow-band imaging surveys \citep[e.g.,][]{ly11,jlee12}, and similar strategies can be adopted to identify and characterize sources with PAH emission.
Sources that show a significant $\rm F300M-F335M$ color excess (i.e., a continuum-subtracted F335M flux) are identified as PAH emitters.  The locus of points around $\rm F300M-F335M=0$ represents continuum sources with little-to-no PAH emission. These objects are typically associated with star clusters older than 10 Myr as shown in panel {\it c} of Fig.~\ref{fig:cmd}. 
Not surprisingly, this distribution becomes broader at fainter F335M magnitudes, due to increasing noise in the $\rm F300M-F335M$ measurement \citep[compare with Fig.~3 in][]{jlee12}. Sources with PAH emission show significant $\rm F300M-F335M$ color excess and are found to the right of the locus, and we find that all 12 prototypes are in this region.  

Based on this clear separation in the CMD, we select a larger sample of bright PAH emitters for examination, i.e., $\rm F300M-F335M>0.3$, $\rm F335M<24$ and $\rm F300M<25$. This selection provides a sample of 67 objects, shown with cyan symbols in panel {\it a} of Fig.~\ref{fig:cmd}. In this figure, the lines indicate the limits $\rm F300M-F335M>0.3$ and $F335M<24$. We note there are additional sources (beyond the 67 cyan candidates) that fall inside this box; however these sources are fainter than $F300M = 25$~mag and not included in our sample.

We visually inspect each of these 67 cluster candidates, and divide them into 3 categories: 
\begin{itemize}
   \item embedded: optical emission very faint or undetected, similar to the 12 prototype objects ($N=28$)
    \item intermediate/ambiguous: optical emission detected, but it is not clear whether the optical source is the counterpart to the F335M source due to positional differences, and lower resolution at 3$\mu$m ($N=16$)
    \item visible/exposed: clear optical counterpart ($N=23$)
\end{itemize}
Examples of objects in each category are shown in Figure~\ref{fig:galaxy+reg}.  We will compare our new, dust embedded cluster sample with the optically selected PHANGS-HST catalog in the Discussion.

\begin{figure*}
\centering
\includegraphics[width=2\columnwidth]{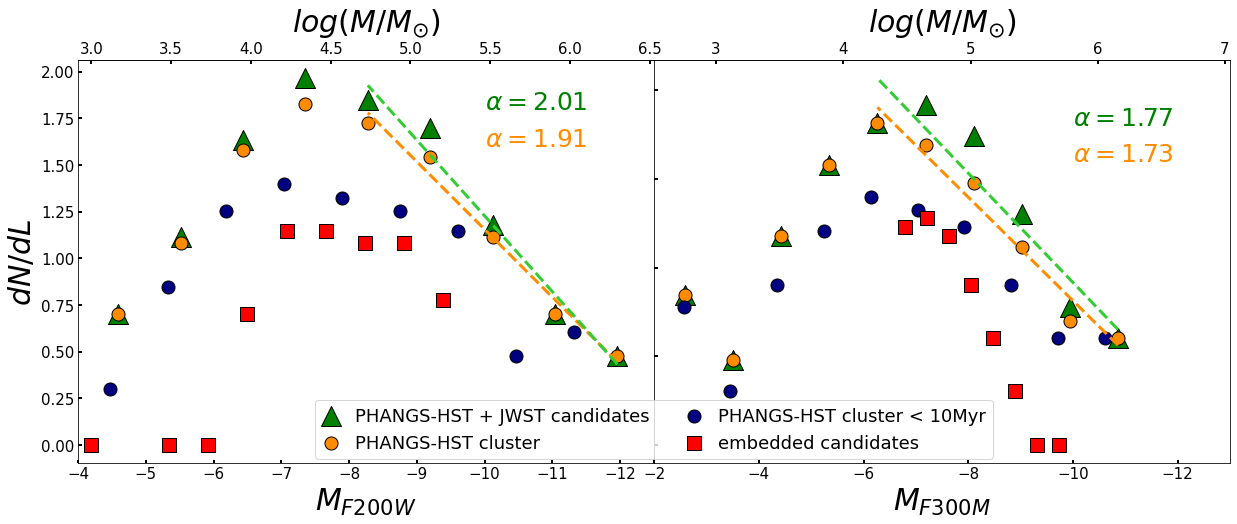}
\caption{Luminosity functions measured for different cluster samples are shown in the $\rm F200W$ {\it(left)} and $\rm F300M$ {\it(right)} filters. 
The red squares show distributions for the 67 dust embedded cluster candidates from this work.
The orange points show the luminosity functions for PHANGS-HST selected clusters at all ages (N=263), and the blue dots are PHANGS-HST clusters younger than 10~Myr (N=131). 
Green triangles indicate the LF of PHANGS-HST clusters plus the 67 embedded cluster candidates (N=330).
The green line shows the best fit index to this last sample, while the orange line show the fit for only the PHANGS-HST cluster distribution (orange circles). 
Our dust embedded cluster sample (red squares) is not sufficiently deep to have many data points in the luminosity function before the distribution begins to flatten, likely due to incompleteness, but comparing the derived slopes with and without the embedded candidates we observe that they are quite similar.
The bright portion of each distribution appears to follow a trend that is consistent with that observed for the optically-selected HST clusters. The upper axis indicates mass estimates from the method described in Section~4.2.}
\label{Fig:LF}
\end{figure*}

\section{Results} \label{subsec:Results}

Luminosity and mass functions for the youngest cluster populations in galaxies give direct constraints on the initial cluster mass function, analogous to the stellar initial stellar mass function.  One of the major weaknesses in current studies of star formation is that the fraction of the most recently formed dust obscured clusters has been unknown and unaccounted for.  These initial results provide a preview of the work with JWST which will measure these fundamental distributions.

\subsection{Luminosity Function}

For clusters with similar ages and reddening, the luminosity function
of clusters can serve as a proxy for their mass function.  The
luminosity function (LF) can be approximated, at least to first order,
by a power law shape of the form $dN/dL \propto L^{\alpha}$
\citep{whitmore99,fall06,whitmore14a}. Values of the power-law index
$\alpha \sim -2$ are typical for young star clusters
\citep[e.g.][]{larsen09,whitmore14a}. 

In Fig~\ref{Fig:LF} we present
the LFs of our embedded cluster candidates in the $\rm F200W$
and $\rm F300M$ bands, which sample flux from the stellar continuum, with some contribution from the dust continuum (red squares).  These distributions increase starting from the brightest magnitudes (to the right), eventually flatten, and then drop off again (F200W filter).  The flattening towards fainter magnitudes is almost certainly due to incompleteness.

 For comparison we also show the LF of optically-selected clusters from the PHANGS-HST survey \citep{phangs-hst,thilker21,turner21,deger22}, with no restriction on age (shown in orange) and clusters with best fit ages younger than 10~Myr (shown in blue). We also plot the LF of the combined cluster sample, i.e. PHANGS-HST clusters plus the 67 embedded candidates (shown with green triangles). The LF for the full HST+JWST cluster sample (green) has a best fit power-law index of $-2.01$ in $\rm F200W$ and $-1.77$ for $\rm F300M$ (shown as the green lines in each panel). The fit over only the PHANGS-HST clusters (orange) gives very similar index values,  $-1.91$ in $\rm F200W$ and $-1.73$ for $\rm F300M$ (orange lines). On the other hand,
 while there are only few data points before our embedded cluster distribution  (red points) flattens at magnitudes fainter than M$_{F200W} \geq$-8, the bright points suggest a behavior similar to the optically-selected clusters. 
The LF in the F300M filter however, appears to be steeper than the PHANGS-HST clusters. This difference could be due to the imposed selection criteria that involve cuts using the F300M filter (Sect.~\ref{subsect:selectioncriteria}).

Based on our preliminary results, we suspect that the LF for embedded star clusters present a similar shape to that for optical clusters. However our sample is not large enough to derive accurate power-law indices.
In future work we will extend this work to the full sample of 19 PHANGS-JWST galaxies, which will be based on total fluxes computed with appropriate aperture corrections.
A larger sample will better constrain the shape of the LF as well as the embedded cluster mass function, which again, is effectively the initial cluster mass function.  Comparison with what we have already learned for visible clusters, will provide important insights into the survival of clusters in transition to becoming exposed/visible. 

\subsection{Estimate of Cluster Masses}

To obtain initial stellar mass estimates for our 67 embedded cluster candidates, we fit a linear relation to obtain the mass-to-light ratio \citep[e.g.][]{fall06,larsen09} using clusters younger than 2~Myr in the PHANGS-HST catalog. In doing so, we obtain coefficients $\gamma$ and $C$, where: $\mbox{log}(M/M_{\odot})=\gamma~\mbox{mag} +C$. We use apparent magnitudes in F200W and F300M bands, obtaining values of $\gamma=$-2.35, $C=$34.22, and $\gamma=$-2.43, $C=$35.46 respectively. 

Using the relation for $\rm F200W$, we obtain a range of masses between 6.4$\times 10^2$ and $1.1 \times 10^5M_{\odot}$, with approximately $80\%$ in the range $10^4$-- $10^5~M_{\odot}$. In the case of masses obtained from $\rm F300M$, the values for our candidates start at $\sim 2\times 10^{4}M_{\odot}$ due to our imposed condition in the selection criteria ($\rm F300M<25$, see Sect.~\ref{subsect:selectioncriteria}). We find that $\sim 88\%$ of the sample is between this minimum value and $10^5~M_{\odot}$, with a few candidates reaching up $\sim 5~\times 10^5M_{\odot}$. The masses estimated from F200W and F300M are generally consistent to within a factor of a few, but clearly, from the vastly different values on the low end of the range, there are candidates (n$\sim$10) with estimates which differ by up to a factor of $\sim$30.  All of these candidates are embedded sources, which likely suffer from high extinction, even at 2$\mu$m.  In future work, we will provide more reliable mass estimates from spectral energy distribution (SED) fitting, as discussed further in the next section. 

\section{Discussion} \label{subsec:Discussion}

We compare our sample of 67 dust embedded cluster candidates with the PHANGS-HST human inspected class~$1+2$ cluster catalog (N=263), and {\em find that our analysis of the PHANGS-JWST imaging has identified 59 new, embedded clusters candidates in NGC~7496.}  These objects are of particular importance because they are likely to be among the youngest clusters in this galaxy, and are needed to construct a complete census.  In this context, it is notable that the size of our new sample is comparable to the number of young HST clusters (59 younger than 2 Myr; 94 younger than 5 Myr), and may double the census of the population younger than $\sim$2 Myr.



 Of the 8 clusters in common between the two samples, we classify seven as visible/exposed, and 1 as partially embedded/intermediate. 
All are found to be younger than 11 Myr (six having ages less than 1~Myr), with masses of $\sim 1-7 \times 10^4 M_{\odot}$.  (As mentioned earlier the full PHANGS-HST cluster sample, meanwhile, contains clusters as old as 13 Gyr.)

We can gain additional insight by examining the infrared properties of clusters of all ages from the optically-selected PHANGS-HST sample.  
Panels {\it c)} and {\it d)} in Fig.~\ref{fig:cmd} show the location of clusters in different age intervals in the F300M-F335M color vs. F335M magnitude diagram.
In panel {\it c)} we can see that clusters older than 100~Myr (magenta triangles) have $\rm F300M-F335M=0$, indicating these older sources do not have PAH emission. This panel also shows that nearly all clusters with ages between 10-100 Myr (green crosses) and 5-10 Myr (yellow diamonds) have $\rm F300M-F335M=0$.
On the other hand, panel {\it d)} shows that all HST clusters with a $\rm F300M-F335M$ color excess (i.e. strong PAH emission) have ages $\leq$2~Myr (cyan crosses). These very young, optically-selected clusters have a similar range of $\rm F300M-F335M$ values as our 67 embedded candidates, although most of them have a significantly fainter F335M magnitude. 
Panel {\it d)} of Fig.~\ref{fig:cmd} also shows clusters with estimated ages between 2 and 5~Myr (orange squares).  While a small fraction of these 2-5~Myr cluster have slighly higher  $\rm F300M-F335M$ values than clusters older than 5~Myr, a surprising number of them show little-to-no PAH emission.  Assuming that the age-dating from PHANGS-HST is correct for these clusters, this result indicates that measurable PAH emission may drop very rapidly (in 2~Myr) in clusters of moderate mass.

\begin{figure}
\centering
\includegraphics[width=\columnwidth]{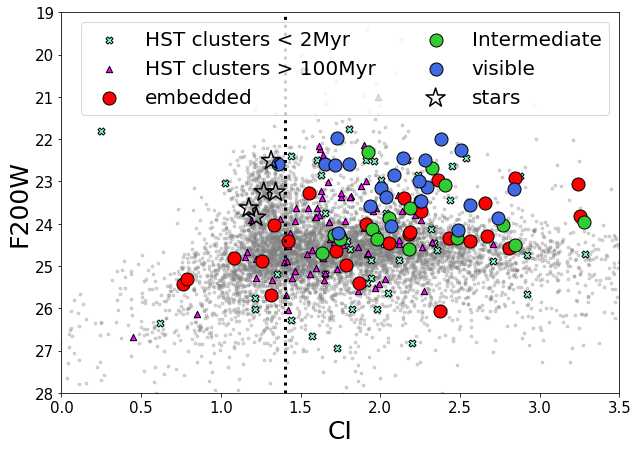}
\caption{F200W magnitude vs. Concentration Index (measured as the difference in magnitude between circular apertures with 1 and 4 pixel radius). The vertical dashed line indicates CI$=1.4$ which can be used to distinguish clusters from stars (Whitmore et al. 2022, this volume). The colored circles represent our 67 cluster candidates as indicated, stars/point sources are represented by a star symbol. PHANGS-HST clusters younger than 2~Myr are small cyan symbols and those older than 100~Myr are small magenta triangles. The grey dots are all the sources detected in the F335M image.}
\label{Fig:CI}
\end{figure}

We also examine the Concentration Index (CI) of infrared and optically selected clusters. The CI provides a measure of the compactness of the source, and is useful to distinguish between clusters and stars \citep[e.g.][]{chandar10b, whitmore14, deger22}. In particular, Whitmore et al. (2022, this volume) found that CI estimated from the JWST F200W filter significantly improves the separation of stars and clusters. CI values were estimated following the procedure in Whitmore et al. (2022, this volume), as the magnitude difference measured between 1 and 4~pixel apertures  (0\farcs031 and 0\farcs124) in the F200W filter.
In Fig~\ref{Fig:CI} we present the results. The vertical dashed line indicates CI$=1.4$, a limit to separate clusters from stars (Whitmore et al. 2022, this volume). The 67 dust embedded clusters from this work are nearly all to the right of the line, meaning they are extended relative to a point source, although there are a few fainter objects which fall to the left of the line.  The dispersion in the size measurements clearly increases towards fainter magnitudes.

\begin{figure*}
\centering
\includegraphics[width=2\columnwidth]{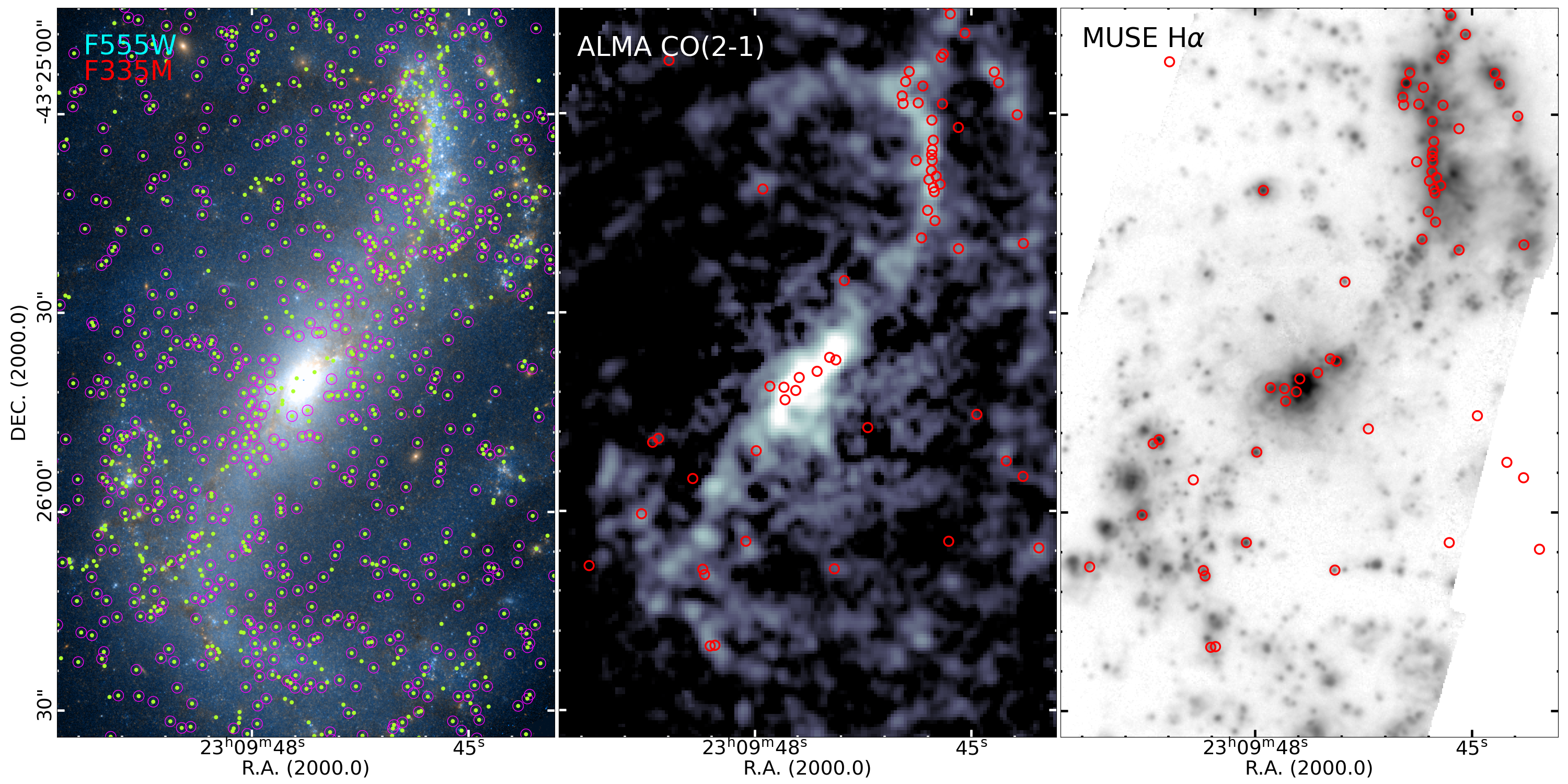}
\caption{{\it Left:} Combined HST F555W (blue) and JWST F335M (red) image. The green dots are sources detected in F335M that satisfy our color cut $\rm F300M-F335M>0.3$ (Sect.~\ref{subsect:selectioncriteria}), but before performing any cut in magnitude.  Sources shown in green without a magenta circle are the strongest PAH emitters, and are mostly found in the main dust lanes within the spiral arms.
Sources with an additional magenta circle are those that have $\rm F300M-F335M$ between 0.3 and 0.7, and $\rm F335M>24.5$, i.e., they are both fainter and have larger color uncertainties (magenta box in Fig.~\ref{fig:cmd}). This figure shows that when we select only the brightest sources with the largest color excesses, sources which tend to coincide with minor dust lanes in the galaxy, i.e., potentially fainter embedded clusters, are excluded.  
{\it Middle}: CO(2-1) ALMA map, with resolution of $\sim 1"$ with the 67 embedded cluster candidates (red circles).
{\it Right}: Muse H$_{\alpha}$ map with resolution of $\sim 0\farcs$7 and the 67 candidates. Note that the sample is well-correlated with $H_{\alpha}$ and CO(2-1) peaks. }
\label{Fig:color_CO_alpha}
\end{figure*}

Next we examine the overall spatial distribution of the sample, as well as the location of our dust embedded cluster candidates relative to other structures and components of the ISM.
In Fig.~\ref{fig:galaxy+reg} we see a strong concentration of embedded clusters (shown in red) along a dust lane in the northern spiral arm, which forms a ridge just next to young (age $<$ 10~Myr) optically-selected clusters. Jumping to the left panel of Fig.~\ref{Fig:color_CO_alpha}, we see that strongest PAH emitters (shown as the green dots, without an accompanying magenta circle) are mostly located where we see the most prominent dust features in the optical HST images, along the spiral arms.
Sources with $\rm F300M-F335M > $ 0.3, but with faint $\rm F335M$ magnitudes (see magenta box in Fig.~\ref{fig:cmd}~{\it a}) are enclosed with a magenta circle.  
Recall that we decided to apply cuts in magnitude to study only the brightest sources ($\rm F335M < 24$ and $\rm F300M < 25$) to simplify this initial analysis.  However, by imposing these limits, weaker PAH emitters are excluded from inspection of the image, these appear to coincide with the more subtle dust lanes and potentially may also be dust embedded clusters.  In future work we will test selection criteria which employ a color excess curve to identify fainter embedded clusters.

The middle and right panels of Fig.~\ref{Fig:color_CO_alpha} show a CO(2-1) ALMA image taken as part of the PHANGS–ALMA survey \citep{leroy21}, and the H$\alpha$ VLT-MUSE image of NGC~7496 from the PHANGS-MUSE survey \citep{emsellem2022}. Our 67 PAH-emitting sources are closely associated with molecular gas and current star formation.
Only 9\% of our candidates appear unassociated with significant peaks of CO emission, H$_{\alpha}$, or both. 
These 9\%  all have no obvious optical counterpart in the HST images, and half have CI values similar to stars, indicating they are very compact and possibly not embedded clusters.

\begin{figure}
\centering
\includegraphics[width=\columnwidth]{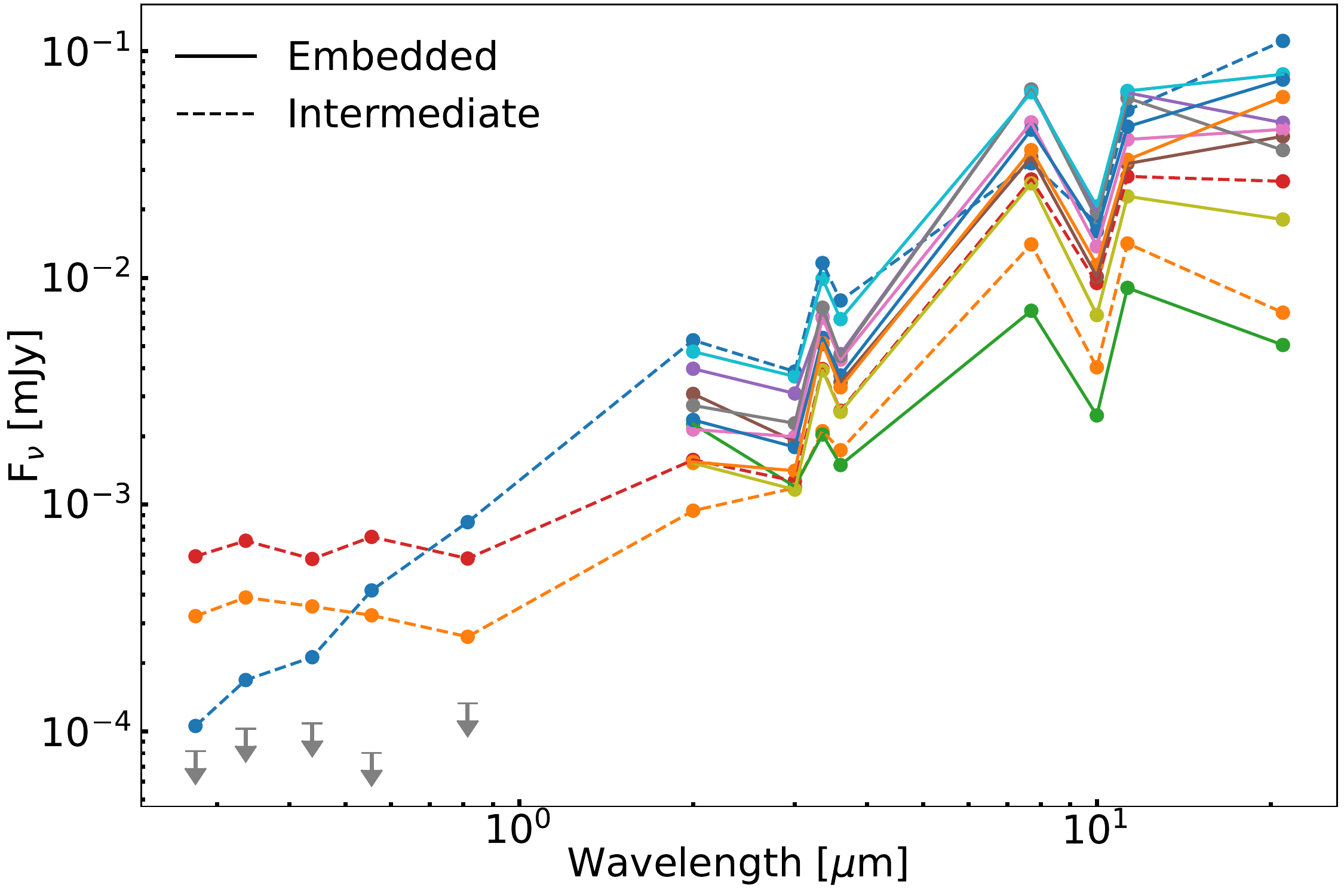}
\caption{ Preliminary UV-IR Spectral Energy Distribution of the 12 embedded prototype clusters (Sect.~\ref{subsec:prototypes}). 
Prototypes classified as partially embedded or `intermediate' are shown with dashed lines, and embedded clusters as solid lines. For objects which are not detected in the HST-bands, we show upper limits with grey arrows. These preliminary SEDs are constructed with fluxes measured with the same aperture in all bands, which leads to an underestimate of the flux in the MIRI-bands, and without aperture correction.
Nevertheless, two features in the SEDs of embedded clusters are clearly seen: PAH emission at $3.3\,\mu$m and silicate absorption at $10\,\mu$m.}
\label{Fig:SED}
\end{figure}

Finally, we examine the SEDs of the embedded clusters from optical wavelengths with HST to the IR with NIRCam and MIRI, which provide further insights into their properties. We show preliminary SEDs of the 12 embedded cluster prototypes in Fig.~\ref{Fig:SED}. These show a clear deficit in optical flux relative to the infrared. In fact, we were able to identify possibly associated faint optical emission for only three of these objects.
For the 9 embedded clusters, without UV-optical detections, we show upper limits in the HST bands. For this first look, the photometric measurements in all bands have been performed with the same 0\farcs155 aperture, which leads to a relative underestimate of the measured fluxes in the MIRI-bands, as no aperture corrections have been applied as discussed before.
To produce proper SEDs for these clusters in future work, we will need to correctly distinguish between crowded point sources and their corresponding diffuse PAH emission, with more careful source extraction which takes the different PSFs into account. These are some of the challenge that need to be addressed in order to perform proper SED fitting for embedded clusters and more accurately constrain their physical properties (e.g., stellar masses, dust masses, ages, reddenings). In addition to the distinctive PAH-feature at $3.3\,\mu$m, we see a dip at $10\,\mu$m indicating silicate absorption, which is caused by protostellar envelopes \citep{2008A&A...486..245C} and has been observed in young stellar objects \citep[e.g.][]{2019A&A...622A.149G}. This spectral feature supports our conclusion that these candidates are very young embedded clusters.


\section{Summary and Conclusions} \label{subsec:Conclusions}

In this work, we have used new NIRCam imaging of the nearby spiral galaxy NGC~7496, the first galaxy to be observed for the PHANGS-JWST Treasury survey, to identify very young clusters embedded in natal dust and gas.  Our selection method relies on detection of PAH 3.3~$\mu$m emission, which is closely associated with the ISM of very recently formed clusters.

We begin by identifying 12 prototype, dust-obscured clusters based on visual inspection of both PHANGS-JWST NIRCam and PHANGS-HST optical images.  We find that all 12 show strong PAH emission at 3.3~$\mu$m, and fall in a region of the F335M vs. $\rm F330M - F335M$ color-magnitude diagram that clearly separates from the vast majority of detections.  
To first order the PAH emission at $3.3\,\mu$m appears to correlate with other dust emission (Sandstrom et al., Dale et al., Chastenet et al., 2022 this volume) and so should represent a sensitive, high resolution tracer of dust-reprocessed UV light from young stars. 
Based on this, we develop criteria to select a larger sample of candidates.  This selection is primarily based on the F330M-F335M color excess, and yields a sample of 67 candidate embedded clusters in NGC~7496.  By comparing the F300M-F335M colors of this sample to optically-selected clusters in the PHANGS-HST catalog, which have ages determined through SED fitting of 5-band UV-optical photometry, we estimate that the embedded clusters are likely to be younger than $\sim$2 Myr.


We find that our new sample may significantly increase the census of young clusters from the PHANGS-HST catalog -- the number of clusters younger than 2 Myr could be increased by a factor of two.  A more careful analysis of the selection functions and completeness of both the HST and JWST cluster samples will be required to provide an accurate assessment of the fraction of young clusters which are embedded and are missing from optical catalogs.  The eight clusters in common between the two catalogs have very young SED ages (six are $\sim1$~Myr).  These results underscore the importance of our new PHANGS-JWST imaging for studying the earliest stages of star and cluster formation. 

Preliminary optical/infrared SEDs for the visually selected prototype candidates show PAH emission and possible silicate absorption. Concentration index measurements confirm that the majority of our 67 sources are more extended than stars/point sources. These characteristics provide additional evidence that most of our selected candidates are star clusters.  We also take a first look at the 2 and 3 $\mu$m luminosity functions for our sample, and estimate stellar masses, finding values of $10^4$-- $10^5~M_{\odot}$.

The strongest and brightest PAH emitters are mainly located in prominent dust lanes within the main spiral arms of the galaxy, next to young optically-detected clusters from PHANGS-HST. The spatial distribution of our candidates show that they are also correlated with H$_{\alpha}$ and CO(2-1) emission. 

In future work we will expand this analysis to all 19 galaxies in the PHANGS-JWST sample, and examine the completeness of samples based on the F300M-F335M color excess for selecting young embedded star clusters.  Significant effort will be required to develop techniques to obtain robust total fluxes across the 13 PHANGS-HST and JWST filters from 2750\AA~to 2$\mu$m due to the factor of 10 decrease in resolution.  Such techniques are needed to produce SEDs appropriate for constraining the physical properties of these clusters through the fitting of stellar population and dust models.  Ultimately, a complete census of star clusters from PHANGS-HST and JWST will enable basic unsolved problems in star formation and cluster evolution to be addressed through new constraints on the initial cluster mass function, star cluster formation efficiencies, as well as on feedback timescales associated with the clearing of the natal gas and dust.  

\section*{acknowledgments}
This work is based on observations made with the NASA/ESA/CSA JWST and NASA/ESA Hubble Space Telescopes. The data were obtained from the Mikulski Archive for Space Telescopes at the Space Telescope Science Institute, which is operated by the Association of Universities for Research in Astronomy, Inc., under NASA contract NAS 5-03127 for JWST and NASA contract NAS 5-26555 for HST. The JWST observations are associated with program 2107, and those from HST with program 15454. The specific JWST observations analyzed can be accessed via \dataset[10.17909/9bdf-jn24]{http://dx.doi.org/10.17909/9bdf-jn24}.  We have also used higher level data products developed and released by PHANGS-HST \dataset[10.17909/t9-r08f-dq31]{http://dx.doi.org/10.17909/t9-r08f-dq31} (image products) \dataset[10.17909/jray-9798]{http://dx.doi.org/10.17909/jray-9798} (catalog products).  We are grateful to Scott Fleming at STScI and his team for their excellent support in helping us provide these products to the community through \url{https://archive.stsci.edu/hlsp/phangs-hst} and \url{https://archive.stsci.edu/hlsp/phangs-cat.}
Based on observations collected at the European Southern Observatory under ESO programmes 1100.B-0651 (PHANGS-MUSE; PI: Schinnerer).
This paper makes use of the following ALMA data: \linebreak
ADS/JAO.ALMA\#2017.1.00886.L, 
ALMA is a partnership of ESO (representing its member states), NSF (USA) and NINS (Japan), together with NRC (Canada), MOST and ASIAA (Taiwan), and KASI (Republic of Korea), in cooperation with the Republic of Chile. The Joint ALMA Observatory is operated by ESO, AUI/NRAO and NAOJ.
JMDK gratefully acknowledges funding from the European Research Council (ERC) under the European Union's Horizon 2020 research and innovation programme via the ERC Starting Grant MUSTANG (grant agreement number 714907). 
COOL Research DAO is a Decentralized Autonomous Organization supporting research in astrophysics aimed at uncovering our cosmic origins.
EJW acknowledges the funding provided by the Deutsche Forschungsgemeinschaft (DFG, German Research Foundation) -- Project-ID 138713538 -- SFB 881 (``The Milky Way System'', subproject P1).
MB acknowledges support from FONDECYT regular grant 1211000 and by the ANID BASAL project FB210003.
TGW acknowledges funding from the European Research Council (ERC) under the European Union’s Horizon 2020 research and innovation programme (grant agreement No. 694343).
JK gratefully acknowledges funding from the Deutsche Forschungsgemeinschaft (DFG, German Research Foundation) through the DFG Sachbeihilfe (grant number KR4801/2-1).
MC gratefully acknowledges funding from the DFG through an Emmy Noether Research Group (grant number CH2137/1-1).
FB would like to acknowledge funding from the European Research Council (ERC) under the European Union’s Horizon 2020 research and innovation programme (grant agreement No.726384/Empire).
RSK acknowledges financial support from the European Research Council via the ERC Synergy Grant ``ECOGAL'' (project ID 855130), from the Deutsche Forschungsgemeinschaft (DFG) via the Collaborative Research Center ``The Milky Way System''  (SFB 881 -- funding ID 138713538 -- subprojects A1, B1, B2 and B8) and from the Heidelberg Cluster of Excellence (EXC 2181 - 390900948) ``STRUCTURES'', funded by the German Excellence Strategy. RSK also thanks the German Ministry for Economic Affairs and Climate Action for funding in the project ``MAINN'' (funding ID 50OO2206). 
ER acknowledges the support of the Natural Sciences and Engineering Research Council of Canada (NSERC), funding reference number RGPIN-2022-03499.
KG is supported by the Australian Research Council through the Discovery Early Career Researcher Award (DECRA) Fellowship DE220100766 funded by the Australian Government. 
KG is supported by the Australian Research Council Centre of Excellence for All Sky Astrophysics in 3 Dimensions (ASTRO~3D), through project number CE170100013. 
SD is supported by funding from the European Research Council (ERC) under the European Union’s Horizon 2020 research and innovation programme (grant agreement no. 101018897 CosmicExplorer).
KK gratefully acknowledges funding from the Deutsche Forschungsgemeinschaft (DFG, German Research Foundation) in the form of an Emmy Noether Research Group (grant number KR4598/2-1, PI Kreckel).
HH acknowledges the support of the Natural Sciences and Engineering Research Council of Canada (NSERC), funding reference number RGPIN-2022-03499.

JPe acknowledges support by the DAOISM grant ANR-21-CE31-0010 and by the Programme National ``Physique et Chimie du Milieu Interstellaire'' (PCMI) of CNRS/INSU with INC/INP, co-funded by CEA and CNES.
G.A.B. acknowledges the support from ANID Basal project FB210003.


%

\vspace{5mm}
\facilities{JWST, HST, ALMA, MUSE}


\software{Astropy \citep{astropy:2013, astropy:2018, astropy:2022}.
This research made use of Photutils, an Astropy package for
detection and photometry of astronomical sources \citep{larry_bradley_2022_6825092}
AstroPy (astropy.org)
          }



\bibliographystyle{aasjournal}
\bibliography{all} 



\end{document}